\documentclass[12pt]{article}
\usepackage[hypertex]{hyperref}
\usepackage[dvipdfmx]{graphicx} 
\usepackage{graphicx,amsmath,amssymb,amsfonts,cite,bm}

    \def\d{\delta}  \def\e{\epsilon} \def\z{\zeta} \def\h{\eta} \def\th{\theta}   \def\l{\lambda}  \def\m{\mu} \def\n{\nu}      \def\s{\sigma}  \def\t{\tau}     

\def\dg{\dagger}  \def\nn{\nonumber}

\newcommand{\lsp}{ \left ( }
\newcommand{\rsp}{ \right ) }

\newcommand{\To}{\Rightarrow}

\newcommand{\vev}[1]{ \langle {#1} \rangle }
\newcommand{\meV}{ {\rm meV} }
\newcommand{\eV}{ {\rm eV} }

\newcommand{\MeV}{ {\rm MeV} }

\def\ds{\displaystyle}

\newcommand{\Column}[3]{ \begin{pmatrix}  #1 \\ #2 \\ #3  \end{pmatrix}  }

\newcommand{\Diag}[3]{ \begin{pmatrix}  #1 & 0 & 0 \\ 0 & #2 & 0 \\ 0 & 0 & #3 \\ \end{pmatrix} }

\newcommand{\democratic}{
\begin{pmatrix}
 1 & 1 & 1 \\ 1 & 1 & 1 \\ 1 & 1 & 1 \\
\end{pmatrix}
}

\newcommand{\DMC}{
\begin{pmatrix}
\frac{1}{\sqrt{2}} & \frac{1}{\sqrt{6}} &  \frac{1}{\sqrt{3}} \\
-\frac{1}{\sqrt{2}} & \frac{1}{\sqrt{6}} &  \frac{1}{\sqrt{3}} \\
0 & -\frac{\sqrt{2}}{\sqrt{3}}  & \frac{1}{\sqrt{3}} 
 \end{pmatrix}
}

\newcommand{\DMCdg}{
\begin{pmatrix}
 \frac{1}{\sqrt{2}} & -\frac{1}{\sqrt{2}} & 0 \\
 \frac{1}{\sqrt{6}} & \frac{1}{\sqrt{6}} & -\frac{\sqrt{2}}{\sqrt{3}}  \\
  \frac{1}{\sqrt{3}} & \frac{1}{\sqrt{3}} & \frac{1}{\sqrt{3}} \\
 \end{pmatrix}
}

\begin{document}

\begin{titlepage}

\begin{flushright}
STUPP-16-227
\end{flushright}

\vskip 1.35cm

\begin{center}
{\LARGE \bf Hierarchical majorana neutrinos \\ from democratic mass matrices}

\vskip 1.2cm

Masaki J. S. Yang

\vskip 0.4cm

{\it Department of Physics, Saitama University, \\
Shimo-okubo, Sakura-ku, Saitama, 338-8570, Japan\\
}


\begin{abstract} 

In this paper, we obtain the light neutrino masses and mixings consistent with the experiments, 
in the democratic texture approach. 
The essential ansatz is that $\n_{Ri}$ are assumed to transform as ``right-handed fields'' $\bf 2_{R} + 1_{R}$ under the $S_{3L} \times S_{3R}$ symmetry.  
The symmetry breaking terms are assumed to be diagonal and hierarchical. 
This setup only allows the normal hierarchy of the neutrino mass, 
and excludes both of inverted hierarchical and degenerated neutrinos. 

Although the neutrino sector has nine free parameters, 
several predictions are obtained at the leading order. 
When we neglect the smallest parameters $\z_{\n}$ and $\z_{R}$, 
all components of the mixing matrix $U_{\rm PMNS}$ are expressed by 
the masses of light neutrinos and charged leptons. 
 From the consistency between predicted and observed $U_{\rm PMNS}$, 
 we obtain the lightest neutrino masses $m_{1}$ = (1.1 $\to$ 1.4)  meV, 
 and the effective mass for the double beta decay $\vev{m_{ee}}\simeq$ 4.5 meV. 

\end{abstract} 

\end{center}
\end{titlepage}

\section{Introduction}

The observation of the neutrino oscillation \cite{Fukuda:1998mi,Ahmad:2002jz} 
clarified finite masses of the neutrinos  and lepton flavor nonconservation.
Furthermore, the Daya Bay and RENO experiments \cite{An:2012eh, Ahn:2012nd} discovered that $U_{e3}$  is nonzero and relatively large. 
However, these experiments shed us a further mysteries, {\it e.g.,} dozen of unexplained parameters, and the origin of the flavor.
In particular, the lepton mixing matrix $U_{\rm PMNS}$  \cite{Maki:1962mu, Pontecorvo:1967fh} is remarkably different from the quark mixing matrix $U_{\rm CKM}$ \cite{Cabibbo:1963yz, Kobayashi:1973fv}.

Innumerable models has been proposed so far, to explain the mysterious flavor structures of the standard model. 
As representative approaches, researchers explore the continuous or discrete flavor symmetries \cite{Froggatt:1978nt, Ishimori:2010au, Varzielas:2012as}, 
and specific flavor textures \cite{Fritzsch:1977vd, Fritzsch:1999ee}.
In the texture approach, the democratic texture \cite{Harari:1978yi, Koide:1983qe, Koide:1989zt, Tanimoto:1989qh, Fritzsch:1989qm, Fritzsch:1995dj, Fukugita:1998vn, Fritzsch:2004xc, Xing:2010iu, Zhou:2011nu}, realized by the $S_{3L} \times S_{3R}$ symmetry is widely studied. 
It assumes that the Yukawa interactions of the fermions $f = u,d,e$ have the ``democratic matrix'' in  Eq.~(\ref{yukawa0}).
In particular, Fujii, Hamaguchi and Yanagida \cite{Fujii:2002jw} has derived the large mixing angles of light neutrinos by the seesaw mechanism \cite{seesaw}, assuming almost degenerated neutrino Yukawa matrix   $Y_{\n} \sim c_{\n} \, {\rm diag} (1,1,1)$. 
This degenerated $Y_{\n}$ is aesthetically unsatisfactory, because it is realized by assuming that the right-handed neutrinos $\n_{Ri}$ transform as ``left-handed fields'' $\bf 2_{L} + 1_{L}$ under the $S_{3L} \times S_{3R}$ symmetry.
Furthermore, the degenerated $Y_{\n}$ is undesirable in viewpoints of grand unified theory (GUT). 
A part of previous authors also have considered the democratic matrices in SU(5) GUT \cite{Fukugita:1998kt}. 
However, degenerated $Y_{\n}$ can not be unified to other Yukawa matrices. 

Then, in this paper, $\n_{Ri}$ are assumed to transform as ``right-handed fields'' $\bf 2_{R} + 1_{R}$ under the $S_{3L} \times S_{3R}$ symmetry.  
The symmetry breaking terms are assumed to be diagonal and hierarchical, 
which is basically same as the previous studies. 
These assumptions realize hierarchical $Y_{\n}$ and forbid degenerated $Y_{\n}$. 
It enables us to treat quarks and leptons uniformly under a simple framework. 
By the seesaw mechanism, we obtain the light neutrino masses 
and mixings consistent with the experiments. 
This setup only allows the normal hierarchy of the neutrino masses, 
and excludes both of inverted hierarchical and degenerated neutrinos. 

Although the neutrino sector has nine free parameters, 
several predictions are obtained at the leading order. 
When we neglect the smallest parameters $\z_{\n}$ and $\z_{R}$, 
the resulting neutrino matrix $m_{\n}$ has only three parameters and then determined from 
the neutrino masses $m_{i}$.  
Therefore, all components of the mixing matrix $U_{\rm PMNS}$ are expressed by 
the masses of light neutrinos and charged leptons. 
 From the consistency between predicted and observed $U_{\rm PMNS}$, 
 we obtain the lightest neutrino masses $m_{1}$ = (1.1 $\to$ 1.4)  meV, 
 and the effective mass for the double beta decay $\vev{m_{ee}}\simeq$ 4.5 meV.  
 
In the second-order perturbation, the predictability becomes little lower. 
However, the hierarchical $Y_{\n}$ can be unified to other Yukawa interactions 
in SO(10) GUT or Pati--Salam models.  
Relating $Y_{\n}$ and $Y_{u}$ in some manner, 
several free parameters in the neutrino sector are expected to be removed. 
Meanwhile, the derivation in this paper remains only at tree level. 
The radiative corrections \cite{Haba:2000rf, Mei:2005gp, Ray:2010rz} and threshold correction \cite{Gupta:2014nba} will modify the results. We leave it for our future work.

This paper is organized as follows. 
In the next section, we review the Yukawa matrices with democratic texture.
In Sect. 3 and 4, we present the parameter analysis of the light neutrino mass.
Section 5 is devoted to conclusions and discussions.

\section{Yukawa matrices with democratic texture}

In the democratic mass matrix approach \cite{Harari:1978yi, Koide:1983qe, Koide:1989zt, Tanimoto:1989qh, Fritzsch:1989qm, Fritzsch:1995dj, Fukugita:1998vn, Fritzsch:2004xc, Xing:2010iu, Zhou:2011nu}, the Yukawa matrices are assumed to be the following texture:
\begin{align}
{Y_{f} = {K_{f} \over 3} \democratic + \Diag{\z_{f}}{\e_{f}}{\d_{f}} ,}
\label{yukawa0}
\end{align}
where $f$ is the SM fermions $f = u,d,e$. 
The first term (often called ``democratic'' mass matrix \cite{Tanimoto:1989qh,Fritzsch:1989qm} is realized by 
assigning fermions $f_{L,R}$ as $\bf 1_{L,R} + 2_{L,R}$ under the $S_{3L} \times S_{3R}$ symmetry;
\begin{align}
f_{(L,R)i}' = S^{(abc)}_{(L,R)ij} f_{(L,R)j}. 
\label{trf}
\end{align}
For example, right-handed fields are explicitly written as
\begin{align}
u_{Ri} = \Column{u_{R}}{c_{R}}{t_{R}}, ~~
d_{Ri} = \Column{d_{R}}{s_{R}}{b_{R}}, ~~
e_{Ri} = \Column{e_{R}}{\m_{R}}{\t_{R}}, ~~
\end{align}
and the left-handed fermions are written as similar way.
The representation of $S^{(abc)}_{ij}$ is 
\begin{eqnarray}
S^{(123)}_{(L,R)} =  \begin{pmatrix} 1 & 0 & 0 \\ 0 & 1 & 0 \\ 0 & 0 & 1 \end{pmatrix} , &
S^{(213)}_{(L,R)} = \begin{pmatrix} 0 & 1 & 0 \\ 1 & 0 & 0 \\ 0 & 0 & 1 \end{pmatrix}  ,  &
S^{(132)}_{(L,R)} = \begin{pmatrix} 1 & 0 & 0 \\ 0 & 0 & 1 \\ 0 & 1 & 0 \end{pmatrix} , \\
S^{(321)}_{(L,R)} = \begin{pmatrix} 0 & 0 & 1 \\ 0 & 1 & 0 \\ 1 & 0 & 0 \end{pmatrix} , &
S^{(312)}_{(L,R)} = \begin{pmatrix} 0 & 0 & 1 \\ 1 & 0 & 0 \\ 0 & 1 & 0 \end{pmatrix} , &
S^{(231)}_{(L,R)} = \begin{pmatrix} 0 & 1 & 0 \\ 0 & 0 & 1 \\ 1 & 0 & 0 \end{pmatrix} .
\end{eqnarray}
The second term in Eq.~(\ref{yukawa0}) breaks the permutation symmetry slightly \cite{Koide:1983qe, Koide:1989zt}. 
Here,  the hierarchical relation 
\begin{align}
K_{f} \gg \d_{f} \gg \e_{f} \gg \z_{f}, 
\label{hier}
\end{align}
is assumed.
For the sake of simplicity of the discussion, we assume all breaking parameters are real. 
The discussion on the CP violation is given later.

The previous study by Fujii, Hamaguchi, and Yanagida \cite{Fujii:2002jw} has derived the large mixing angles of light neutrinos by the seesaw mechanism, assuming almost degenerated neutrino Yukawa matrix   $Y_{\n} \sim c_{\n} \, {\rm diag} (1,1,1)$.
This degenerated $Y_{\n}$ is aesthetically unsatisfactory, because it is realized by assuming that the right-handed neutrinos $\n_{Ri}$ transform as ``left-handed fields'' $\bf 2_{L} + 1_{L}$ under the $S_{3L} \times S_{3R}$ symmetry. Furthermore, the degenerated $Y_{\n}$ is undesirable in viewpoints of grand unified theory (GUT).

Then, in this paper, $\n_{Ri}$ are assumed to transform as ``right-handed fields'' $\bf 2_{R} + 1_{R}$ under the $S_{3L} \times S_{3R}$ symmetry.  
The charge assignment of the leptons are shown in the Table 1. 
The symmetry breaking terms are assumed to be diagonal and hierarchical,
 is basically same as the previous studies. 
These assumptions realize hierarchical $Y_{\n}$ and forbid degenerated $Y_{\n}$.
The texture of Yukawa matrices are determined as Eq.~(\ref{yukawa0}) 
for all SM leptons $f = \n,e$. 
\begin{table}[htb]
  \begin{center}
    \begin{tabular}{|c|c|c|} \hline
          & $S_{3L}$ & $S_{3R}$ \\ \hline \hline
      $l_{L i}$ & $\bf 1_L + 2_L$ & $\bf 1_R$    \\
      $\n_{Ri}, e_{Ri}$ & $\bf 1_L$ &  $\bf 1_R + 2_R$ \\ \hline
    \end{tabular}
    \caption{The charge assignments of the leptons under the discrete symmetries.}
  \end{center}
\end{table}

Due to the charge assignment, the majorana mass term of $\n_{Ri}$ invariant under the $S_{3R}$ symmetry is found to be 
\begin{align}
M_{R} = m_{R} \left [ {K_{R} \over 3} \democratic + c_{R} \Diag{1}{1}{1} + \Diag{\z_{R}}{\e_{R}}{\d_{R}}  \right ] . 
\label{mnr}
\end{align}
Here, we assume the symmetry breaking term to $M_{R}$ is also the diagonal.
The term proportional to $c_{R}$ is forbidden for the $Y_{\n}$ by the assignment.
In order to cancel out the hierarchy of $Y_{\n}$ in the seesaw mechanism, the mass matrix~(\ref{mnr}) should be strongly hierarchical. Then, $K_{R} \gg c_{R}$ is required. 
Since the term $c_{R} \,  {\rm diag} ~ (1,1,1)$ is symmetric under $S_{3L} \times S_{3R}$, the parameter $c_{R}$ need not necessarily be small parameter. Then, we assume 
\begin{align}
K_{R} \gg c_{R} \gg \d_{R} \gg \e_{R} \gg \z_{R} .
\label{hier2}
\end{align}

\vspace{12pt}

When we analyze the matrices~(\ref{yukawa0}), 
at first the democratic matrix is diagonalized by the following unitary matrix:
\begin{align}
U_{\rm DC}^{} = \DMC .
\label{DMC}
\end{align}
It is explicitly written as, 
\begin{align}
&U_{\rm DC}^{\dg} Y_{f} U_{\rm DC}^{} \\
&= 
\DMCdg
\left[ {K_{f} \over 3} \democratic + \Diag{\z_{f}}{\e_{f}}{\d_{f}}  \right] 
\DMC
\\ & =
 K_{f} \Diag{0}{0}{1} + 
\begin{pmatrix}
{1\over 2} (\z_f + \e_f) & {1 \over 2 \sqrt{3}} (\z_f - \e_f) & {1\over \sqrt{6}} (\z_f - \e_f) \\
{1 \over 2 \sqrt{3}} (\z_f - \e_f) & {1\over 6} (\z_f + \e_f + 4 \d_f) &  {1\over 3 \sqrt{2}} (\z_f + \e_f - 2 \d_f) \\
{1\over \sqrt{6}} (\z_f - \e_f) & {1\over 3 \sqrt{2}} (\z_f + \e_f - 2 \d_f) &  {1\over 3} (\z_f + \e_f + \d_f) \\
\end{pmatrix} .
\label{cascade}
\end{align}
From the hierarchical relation~(\ref{hier}), approximate form of this matrix found to be the ``cascade texture'' \cite{Haba:2008dp}
\begin{align}
U_{\rm DC}^{\dg} Y_{f} U_{\rm DC}^{} \cong {1\over 6}
\begin{pmatrix}
3 \e_{f} &  -  \sqrt{3} \e_{f} & - \sqrt{6} \e_{f} \\
- \sqrt{3} \e_{f} & 4  \d_{f} & -  2 \sqrt{2} \d_{f} \\
- \sqrt{6} \e_{f} & - 2 \sqrt{2} \d_{f} &   6 K_{f} \\
\end{pmatrix} .
\label{cascade}
\end{align}
If we assign $\z_{f} = - \e_{f}$, it leads to the zero texture $(U_{\rm DC}^{\dg} Y_{f} U_{\rm DC}^{})_{11} = 0$  \cite{Koide:1983qe, Koide:1989zt, Fritzsch:1995dj,Fritzsch:1998xs}, 
that corresponds the ``hybrid texture'' in Ref.~\cite{Haba:2008dp}.

Eq.~(\ref{cascade}) is perturbatively diagonalized as 
\begin{align}
U_f^{\dg} Y_f U_f = {\rm diag}( y_{1f}, y_{2f}, y_{3f}), 
\end{align}
where
\begin{align}
y_{1f} &=(\z_{f}+\e_{f}+\d_{f})/3-\xi^q/6 , \label{EG1} \\
y_{2f} &=(\z_{f}+\e_{f}+\d_{f})/3+\xi^q/6  ,  \\
y_{3f} &=K_f+(\z_{f}+\e_{f}+\d_{f})/3 ,  
\label{EG3}
\end{align}
with
\begin{align}
\xi_{f} = \sqrt {(2\d_f-\e_f-\z_f)^2+3(\e_f-\z_f)^2 } .
\end{align}
Here, the second order perturbations $O(\z_{f}^{2} / \e_{f}, \e_{f}^{2} / \d_{f}, \d_{f}^{2} / K_{f})$ are all ignored. If we use Eq.~(\ref{hier}), the eigenvalues~(\ref{EG1}) - (\ref{EG3}) are approximated as the diagonal elements of Eq.~(\ref{cascade}),
\begin{align}
y_{1f} \simeq {1\over 2} \e_{f}, ~~~ y_{2f} \simeq {2 \over 3} \d_{f}, ~~~ y_{3f} \simeq K_{f}. 
\end{align}

The unitary matrices $U_f = U_{\rm DC} B_f $ are found to be \cite{Koide:1983qe, Koide:1989zt, Fukugita:1998vn}
\begin{align}
B_f= 
\begin{pmatrix}
\cos \th_{f} & \sin \th_{f} & \l_{f}  \sin 2 \th^{f} \\
- \sin \th_{f} & \cos \th_{f} & - \l_{f}  \cos 2 \th^{f}  \\
- \l_{f} \sin 3 \th^{f} &  \l_{f} \cos 3 \th^{f} & 1
\end{pmatrix} ,
\end{align}
where
\begin{align}
\tan 2\theta_f \simeq-\sqrt{3}{\e_f - \z_f \over 
  2 \d_f - \e_f - \z_f},   ~~~ \l_f= { \xi_f \over 3 \sqrt 2 K_f} .
\label{mixing}
\end{align}
Note that this system can be interpreted as a toy model of the mixing between neutral mesons $\pi^{0}, \h^{0}, \h'{}^{0}$. 
Indeed, in Eq.~(\ref{yukawa0}), the first democratic term corresponds the gluonic anomaly that provide $\h'{}^{0}$ mass and the second term does the small quark masses $m_{u,d, s}$ \cite{Fritzsch:1975tx}. 
The mixing angle $\th_{f}$~(\ref{mixing}) is the same form to the $\pi^{0}$ - $\h^{0}$ mixing in the chiral perturbation theory \cite{Pich:1995bw}. 

The similarity between the Yukawa interactions and the neutral meson mixing is 
indicated since long years ago \cite{Fritzsch:1988ix,Fritzsch:1988hq}, and
it suggests that fermion mass matrices might be ruled by some mass gap phenomena or unknown underlying principle. 

\section{Simplified case: $\z_{\n} = \z_{R} = c_{R} = 0$}

From the Yukawa matrices Eq.~(\ref{yukawa0}) and the mass matrix Eq.~(\ref{mnr}), 
 the small neutrino mass is obtained by the seesaw mechanism \cite{seesaw}
\begin{align}
m_{\n} = {v^{2} \over 2} Y_{\n}^{T} M_{R}^{-1} Y_{\n} ,
\label{seesaw}
\end{align}
where $v/ \sqrt 2 = \vev{H}$ is the vacuum expectation value of the Higgs boson.
As the simple and important example, 
let us consider a simplified parameter set, $\z_{\n} = \z_{R} = c_{R} = 0$.
In this case, the resulting small neutrino mass is also democratic type with the diagonal breaking term:
\begin{align}
m_{\n}^{(0)} = {v^{2} \over 2 } {1 \over m_{R}} \left[ {K^{2}_{\n} \over 3 K_{R}} \democratic +  \Diag{0}{ \e_{\n}^2 / \e _R}{\d_{\nu }^2 / \d _R} \right] .
\label{masslight}
\end{align}
Note that this is the exact results and no approximation is used.

In order to obtain observed large mixing angles of $U_{\rm PMNS}$, 
the diagonalization of $m_{\n}$ should have only small mixing angles. 
Otherwise, the diagonalization of $m_{\n}$ cancel outs that of the charged lepton mass, or $Y_{e}$, 
which is almost diagonalized by $U_{\rm DC}$~(\ref{DMC}).
Then, the following hierarchical relation is required phenomenologically:
\begin{align}
{K^{2}_{\n} \over 3 K_{R}} \ll {\e_{\n}^2 \over \e _R} \ll {\d_{\nu }^2 \over  \d_R} .
\label{hierpheno}
\end{align}
Accordingly, the normal hierarchy (NH) $m_{1} \ll m_{2} \ll m_{3}$ is forced for these parameter sets, 
and both inverted hierarchical and degenerated masses are excluded. 

\vspace{12pt}

If we treat $m_{1}$ as a small perturbation,
the mass matrix~(\ref{masslight}) is diagonalized at the leading order as
\begin{align}
m_{\n}^{(0)} & = 
\begin{pmatrix}
m_{1} & m_{1} & m_{1} \\
m_{1} &  m_{2} & m_{1} \\
m_{1} & m_{1} &  m_{3}
\end{pmatrix}
\equiv V_{\n} m_{\n}^{\rm diag} V_{\n}^{\dg}, 
\\ & \simeq
\begin{pmatrix}
1 & {m_{1} \over m_{2} - m_{1}} & {m_{1} \over m_{3}- m_{1}} \\[5pt]
- {m_{1} \over m_{2} - m_{1}} & 1 & {m_{1} \over m_{3} - m_{1}} \\[5pt]
- {m_{1} \over m_{3} - m_{1}} & -{m_{1} \over m_{3} - m_{1}} & 1\\
\end{pmatrix}
\Diag{m_{1}}{m_{2}}{m_{3}}
\begin{pmatrix}
1 & -{m_{1} \over m_{2} - m_{1}} & -{m_{1} \over m_{3} - m_{1}} \\[5pt]
{m_{1} \over m_{2} - m_{1}} & 1 & -{m_{1} \over m_{3} - m_{1}} \\[5pt]
{m_{1} \over m_{3} - m_{1}} & {m_{1} \over m_{3} - m_{1}} & 1\\
\end{pmatrix} , \label{Vn}
\end{align}
where 
\begin{align}
m_{1} = {v^{2} \over 2 \, m_{R}} {K_{\n}^{2} \over 3 K_{R}}, ~~~
m_{2} = {v^{2} \over 2 \, m_{R}} \lsp {\e_{\n}^{2} \over \e_{R}} + {K_{\n}^{2} \over  3 K_{R}} \rsp , ~~~
m_{3} = {v^{2} \over 2 \, m_{R}} \lsp {\d_{\n}^{2} \over \d_{R}} + {K_{\n}^{2} \over  3 K_{R}} \rsp .
\label{masses}
\end{align}
As a result, the neutrino mixing matrix is calculated as
\begin{align}
U_{\rm PMNS}^{0} &= U_{e}^{\dg} V_{\n} =  B_{e}^{\dg} U_{\rm DC}^{\dg} V_{\n} \\
&\simeq 
\begin{pmatrix}
1 & {m_{e} \over \sqrt{3} m_{\m}} & { 3 m_{e} \over \sqrt 6 m_{\t}} \\[4pt]
- {m_{e} \over \sqrt{3} m_{\m}} & 1 &  {m_{\m} \over \sqrt 2 m_{\t}} \\[4pt]
- { 2 m_{e} \over \sqrt 6 m_{\t}} & -  {m_{\m} \over \sqrt 2 m_{\t}}  & 1 \\
\end{pmatrix} 
\begin{pmatrix}
 \frac{1}{\sqrt{2}} & -\frac{1}{\sqrt{2}} & 0 \\
 \frac{1}{\sqrt{6}} & \frac{1}{\sqrt{6}} & -\frac{\sqrt{2}}{\sqrt{3}}  \\
  \frac{1}{\sqrt{3}} & \frac{1}{\sqrt{3}} & \frac{1}{\sqrt{3}} \\
\end{pmatrix}
\begin{pmatrix}
1 & m_{1} \over m_{2} & m_{1} \over m_{3}  \\[4pt]
- m_{1} \over m_{2}  & 1 & m_{1} \over m_{3}  \\[4pt]
- m_{1} \over m_{3}  & -m_{1} \over m_{3} & 1\\
\end{pmatrix}
\\[8pt] & \simeq
\begin{pmatrix}
\ds \frac{1}{\sqrt{2}} +  \frac{m_{1}}{\sqrt{2} m_{2}} & \ds -\frac{1}{\sqrt{2}} + \frac{m_{1}}{\sqrt{2} m_{2}} & 0 \\
\ds \frac{1}{\sqrt{6}} + \frac{m_{\mu }}{\sqrt{6} m_{\tau }} -\frac{m_{1}}{\sqrt{6} m_{2}} & \ds \frac{1}{\sqrt{6}} +\frac{m_{\mu }}{\sqrt{6} m_{\tau }} +\frac{m_{1}}{\sqrt{6} m_{2}} &\ds -\sqrt{\frac{2}{3}} + \frac{m_{\mu }}{\sqrt{6} m_{\tau }} \\[12pt]
\ds \frac{1}{\sqrt{3}} - \frac{m_{\mu }}{2 \sqrt{3} m_{\tau }}  -\frac{m_{1}}{\sqrt{3} m_{2}} &\ds \frac{1}{\sqrt{3}} - \frac{m_{\mu }}{2 \sqrt{3} m_{\tau }} + \frac{m_{1}}{\sqrt{3} m_{2}} & \ds \frac{1}{\sqrt{3}} + \frac{m_{\mu }}{\sqrt{3} m_{\tau }} \\
\end{pmatrix} .
\label{UPMNS0}
\end{align}
In the final expression, we neglect the ratios $m_{e} / m_{\m}, m_{e}/ m_{\t}$ and $m_{1} / m_{3}$. 

In this simplified case, we have six free parameters. However, at leading order, 
free parameters are only three neutrino masses $m_{i}$ in the mixing matrix (\ref{UPMNS0}) .
 In particular, $U_{\m 3}$ and $U_{\t 3}$ is expressed by masses of heavy leptons:
\begin{align}
U_{\m 3} &\simeq \ds -\sqrt{\frac{2}{3}} + \frac{m_{\mu }}{\sqrt{6} m_{\tau }} \simeq -0.766 , \\
U_{\t 3} &\simeq \ds \frac{1}{\sqrt{3}} + \frac{m_{\mu }}{\sqrt{3} m_{\tau }} \simeq 0.645.
\end{align}
Here, we used the pole masses $m_{\mu }=105.6~\MeV$ and  $m_{\tau }=1776~\MeV$.
These components are in the $3 \s$ range of 
the latest global analysis \cite{Gonzalez-Garcia:2015qrr}:
\begin{align}
| U | = 
\begin{pmatrix}
0.801 \to 0.845 & 0.514 \to 0.580 & 0.137 \to 0.158 \\
0.225 \to 0.517 & 0.441 \to 0.699 & 0.614 \to 0.793 \\
0.246 \to 0.529 & 0.464 \to 0.713 & 0.590 \to 0.776 \\
\end{pmatrix} .
\label{realU}
\end{align}

Note that the difference between the realistic pattern $|U|$~(\ref{realU}) and the leading order mixing $U_{\rm DC}$~(\ref{DMC}) is of $O(0.1)$ for all matrix elements. 
It suggests that $|U|$~(\ref{realU}) might result from properly perturbed $U_{\rm DC}$ \cite{Xing:2012ej}. 
Therefore, in the next section, 
we will explore the proper parameter sets of this neutrino mass system, 
including parameters $\z_{\n}, \z_{R}$ and $c_{R}$.

\section{General case: $\z_{\n} \neq \z_{R} \neq c_{R} \neq 0$}

The system of neutrinos analyzed here has nine free parameters, $ K_{\n, R}, c_{R},  \d_{\n,R}, \e_{\n,R},$ and $ \z_{\n,R}$ ($m_{R}$ is essentially not free parameter because its magnitude can be absorbed into other parameters).
Hereafter, we will treat the following quantities as free parameters: 
\begin{align}
m_{1,2,3}, ~~~ 
c_{R}, ~~~ \z_{\n}, ~~~ \z_{R}, ~~~ 
\frac{K_{\nu }}{K_R} \equiv r_{K}, ~~~ 
\frac{\d_{\n}}{\d_R} \equiv r_{\d}, ~~~ 
\frac{\e _{\nu }}{\e _R} \equiv r_{\e}.
\end{align}
%

\subsection{Case 1: $\z_{\n} \neq \z_{R} \neq 0, ~ c_{R} = 0$}

For the finite (but small) $\z_{\n}, \z_{R}$ and $c_{R} = 0$,  
the mass matrix $m_{\n}$ calculated from the seesaw formula~(\ref{seesaw}) 
will be perturbed expression from the mass matrix for $\z_{\n} = \z_{R} = c_{R}= 0$ (\ref{masslight}).
When we expand the mass matrix in the first order of $\z_{\n}, \z_{R}$, 
the perturbation is found to be
\begin{align}
m^{(1)}_{\n} &= m_{\n}^{(0)} + \delta m_{\n}, 
\end{align}
where 
\begin{align}
 \delta m_{\n} = {v^{2} \over 2 \, m_{R}}
\begin{pmatrix}
 0 & -r_{\e} & -r_{\d} \\
 -r_{\e} & 0 & 0 \\
 -r_{\d} & 0 & 0 \\
\end{pmatrix}
\z_{\nu } 
+ {v^{2} \over 2 \, m_{R}}
\begin{pmatrix}
 0 & 0 & 0 \\
 0 & -r_{\e}^{2} & -r_{\e} r_{\d} \\
 0 & -r_{\e}r_{\d} & -r_{\d}^{2} \\
\end{pmatrix}
\z_R .
\end{align}
Here, we used the relation $r_{\d}, r_{\e} \gg r_{K}$, 
obtained from the hierarchical relations~(\ref{hier}), (\ref{hier2}),
and the phenomenologically required relation~(\ref{hierpheno}).
At first the mass matrix is diagonalized by $V_{\n}^{(0)} = V_{\n}$ in Eq.~(\ref{Vn}), 
and further diagonalized by the proper perturbation $V_{\n}^{(1)}$: 
\begin{align}
m^{(1)}_{\n} &= V_{\n}^{(1) \dg} V_{\n}^{(0) \dg} ( m_{\n}^{(0)} + \delta m_{\n} ) V_{\n}^{(0)} V_{\n}^{(1)}, 
\end{align}
As a result, the mixing matrix is modified from Eq.~(\ref{UPMNS0}) 
\begin{align}
U_{\rm PMNS}^{(1)} &= U_{e}^{\dg} V_{\n}^{(0)} V_{\n}^{(1)}  =  U_{\rm PMNS}^{(0)} V_{\n}^{(1)} .
\end{align}
Although the explicit form of $V_{\n}^{(1)}$ is troublesome, 
$U_{e3}$ found to be
\begin{align}
{U}_{e3} \simeq \frac{\e_{\nu } \z_R}{\sqrt{2} \d_{\n} \e _R}-\frac{\z _{\nu }}{\sqrt{2} \d_{\n}} .
\end{align}
It suggest rather large parameters $\z_{n}, \z_{R} \simeq 0.2 \d_{\n}$, and contradict to the hierarchical  assumption Eqs.~(\ref{hier}), (\ref{hier2}). 
{If we consider the unification between quarks and leptons such as SO(10) GUT, this possibility is undesirable because $Y_{\n} = Y_{u}$ and the resulting $\th_{13}$ is too suppressed by $\z_{\n} / \d_{\n} \lesssim m_{u} / m_{c}, ~ \e_{\n} / \d_{\n} \simeq m_{u} / m_{c}$. }
However, with the assumption $\z_{n}, \z_{R} \simeq 0.2 \d_{\n}$, we found some parameter regions where all elements are in 3 $\s$ range of Eq.~(\ref{realU}).

\subsection{Case 2: $\z_{\n} = \z_{R} = 0, ~ c_{R} \neq 0$}

Since the term $c_{R} \,  {\rm diag} ~ (1,1,1)$ is  symmetric under $S_{3L} \times S_{3R}$ symmetry, 
we assume $c_{R} \gg \d_{R} \gg \e_{R}$, as in Eq.~(\ref{hier2}).
In this case, parameters $\d_{R}, \e_{R}$ does not appear at the leading order. 

The procedure is rather similar to the simplified case.  The mass matrix is found to be
\begin{align}
m_{\n} \simeq {v^{2} \over 2 } {1 \over m_{R}} \left[ {K^{2}_{\n} \over 3 K_{R}} \democratic 
+  { 1 \over 3 c_{R}}
\begin{pmatrix}
0 & 0 & 0 \\
0 & 2 \, \e_{\n}^{2} & - \d_{\n} \e_{\n} \\
0 & - \d_{\n} \e_{\n} &  2 \, \d_{\n}^{2} \\
\end{pmatrix}
\right] .
\label{masslight2}
\end{align}
It is not exact result, because we used hierarchical relation~(\ref{hier2}).

Eq.~(\ref{masslight2}) is diagonalized at the leading order as
\begin{align}
m_{\n} & = 
\begin{pmatrix}
m_{1} & m_{1} & m_{1} \\
m_{1} &  m_{2} & m_{23} \\
m_{1} & m_{23} &  m_{3}
\label{masslight2p}
\end{pmatrix}
\equiv V_{\n} m_{\n}^{\rm diag} V_{\n}^{\dg}, 
\\[6pt]
V_{\n} & \simeq
\begin{pmatrix}
1 & {m_{1} \over m_{2} - m_{1}} & {m_{1} \over m_{3} -m_{1}} \\[4pt]
- {m_{1} \over m_{2} - m_{1}} & 1 & {m_{23} \over m_{3} - m_{2}} \\[4pt]
- {m_{1} \over m_{3} - m_{1}} & -{m_{23} \over m_{3} - m_{2}} & 1\\
\end{pmatrix} . 
\end{align}
Here, 
\begin{align}
m_{1} &= {v^{2} \over 2 \, m_{R}} {K_{\n}^{2} \over 3 K_{R}}, \\
m_{2} &= {v^{2} \over 2 \, m_{R}} \lsp {2 \e_{\n}^{2} \over 3 c_{R}} + {K_{\n}^{2} \over  3 K_{R}} \rsp 
\equiv m_{1} + \d m_{21}, \\
m_{3} &= {v^{2} \over 2 \, m_{R}} \lsp {2 \d_{\n}^{2} \over 3 c_{R}} + {K_{\n}^{2} \over  3 K_{R}} \rsp 
\equiv m_{1} + \d m_{31}.
\label{masses}
\end{align}
and 
\begin{align}
m_{23} = {v^{2} \over 2 \, m_{R}} \lsp {K_{\n}^{2} \over 3 K_{R}} -  {\d_{\n} \e_{\n} \over 3 c_{R}} \rsp. 
\end{align}
Indeed, this parameter $m_{23}$ is not independent from the mass $m_{1}$ and  the mass differences $\d m_{i1} \equiv m_{i} - m_{1} (i = 2,3)$:
\begin{align}
m_{23} = m_{1} - {1\over 2} \sqrt { \d m_{21} \, \d m_{31}} \,  ,
\end{align}
Then, the mass matrix~(\ref{masslight2p}) is determined by the three neutrino masses $m_{i}$. 

In this case, the neutrino mixing matrix $U_{\rm PMNS}$  
is approximately given by product of Eq.~(\ref{UPMNS0}) and a mixing matrix
\begin{align}
U_{\rm PMNS} = B_{e}^{\dg} U_{\rm DC}^{\dg} V_{\n} 
\simeq U_{\rm PMNS}^{(0)} \,
\begin{pmatrix}
1 & 0 & 0 \\
0 & 1 & - m_{23} \over m_{3} - m_{2} \\
0 & m_{23} \over m_{3} - m_{2} & 1
\end{pmatrix} .
\end{align}
Accordingly, $U_{e3}$ is found to be
\begin{align}
U_{e3} \simeq {1\over \sqrt 2} {m_{23} \over m_{3} - m_{2}} \lsp 1 - {m_{1} \over m_{2}} \rsp .
\label{Ue3}
\end{align}
Then, if we treat $U_{e3}$ as a perturbative input parameter, $U_{\rm PMNS}$ is expressed by known parameters except $m_{i}$: 
\begin{align}
U_{\rm PMNS} &\simeq 
\begin{pmatrix}
\frac{1}{\sqrt{2}} +  \frac{m_{1}}{\sqrt{2} m_{2}} & -\frac{1}{\sqrt{2}} + \frac{m_{1}}{\sqrt{2} m_{2}} & 0 \\
\frac{1}{\sqrt{6}} + \frac{m_{\mu }}{\sqrt{6} m_{\tau }} -\frac{m_{1}}{\sqrt{6} m_{2}} & \frac{1}{\sqrt{6}} +\frac{m_{\mu }}{\sqrt{6} m_{\tau }} +\frac{m_{1}}{\sqrt{6} m_{2}} & -\sqrt{\frac{2}{3}} + \frac{m_{\mu }}{\sqrt{6} m_{\tau }} \\[12pt]
\frac{1}{\sqrt{3}} - \frac{m_{\mu }}{2 \sqrt{3} m_{\tau }}  -\frac{m_{1}}{\sqrt{3} m_{2}} &\frac{1}{\sqrt{3}} - \frac{m_{\mu }}{2 \sqrt{3} m_{\tau }} + \frac{m_{1}}{\sqrt{3} m_{2}} & \frac{1}{\sqrt{3}} + \frac{m_{\mu }}{\sqrt{3} m_{\tau }} \\
\end{pmatrix} \nn
\\[10pt] & \times 
\begin{pmatrix}
 1 & 0  & 0 \\[10pt]
0 & 1 & - \sqrt 2 \, U_{e3} \lsp {m_{1} \over m_{2}} -1 \rsp^{-1} \\[10pt]
0 & \sqrt 2 \, U_{e3} \lsp {m_{1} \over m_{2}} -1 \rsp^{-1} & 1 \\
\end{pmatrix} . 
\label{UPMNS1}
\end{align}
At the leading order, $U_{\m3}$ and $U_{\t3}$ are written by all known parameters:
\begin{align}
U_{\m3} &\simeq  -{1\over \sqrt 3} U_{e3} -\sqrt{\frac{2}{3}} + \frac{m_{\mu }}{\sqrt{6} m_{\tau }} = -0.701 \to -0.713, \\
U_{\t 3} &\simeq - \sqrt{2 \over 3} U_{e3} + \frac{1}{\sqrt{3}} + \frac{m_{\mu }}{\sqrt{3} m_{\tau }} = 
0.723 \to 0.740.
\end{align}
Here, we used $m_{\m}=105.6~\MeV, m_{\t}=1776~\MeV,$ and $U_{e3} = -0.137 \to -0.158$ (negative value is more preferred). 
These two elements are in $3 \s$ range of Eq.~(\ref{realU}). 
However, these formula can have roughly $5 \sim 10 \%$ error which come from 
the second order perturbations of $m_{1} / m_{2}, \, m_{2} / m_{3}$, and $U_{e3}$\footnote{The terms like $U_{e3} \, {m_{1} \over m_{2}}$ are also regarded as the second order perturbations.}. They are inevitable predictions of this model (assuming CP conservation). 

If we set $m_{1} = 0$, it determines other masses $\d m_{21} = m_{2} \simeq 0.008 \, \eV$ and
$\d m_{31} = m_{3} \simeq 0.05 \, \eV$. 
 $U_{e3}$ is also determined from Eq.~(\ref{Ue3}),
\begin{align}
U_{e3} = -\frac{\sqrt{m_2 m_3}}{2 \sqrt{2} \left(m_3-m_2\right)} \simeq -0.168.
\end{align}
This value is close to the global fit $|U_{e3}| = 0.137 \to 0.158$. 
Then, treating $m_{1}$ as a perturbative parameter, we can predict $m_{1}$ from the current error of $U_{e3}$: 
\begin{align}
m_{1} = (0.2 \to 0.6) \, \meV .
\label{m1first}
\end{align}

On the other hand, 
consistency between Eq.~(\ref{UPMNS1}) and the latest global analysis Eq.~(\ref{realU}), 
the mass ratio of lighter neutrinos is predicted as
\begin{align}
0.138 \lesssim \, {m_{1} \over m_{2}} \, \lesssim  0.150 .
\end{align}
Then, the mass eigenvalues are found to be
\begin{align}
m_{1} \simeq (1.1  \to 1.4) \, \meV. 
\label{m1second}
\end{align}
There is a tension between the predictions from $U_{e3}$ and $U_{e(1,2,3)}, U_{\m (1,2,3)}$. 
However, if we adopt $\z_{e} = -\e_{e}$ same as the previous study \cite{Fujii:2002jw}, 
$U_{e3}$ becomes finite at the zeroth order, $U_{e3} = - \sqrt{2 m_{e} \over 3 m_{\m}} \simeq -0.056$. 
In this case the tension will be successfully reconciled. 
Then we tentatively discard the prediction~(\ref{m1first}). The relation between $U_{e3}$ and $m_{i}$ will be reevaluated in the next study. 

The neutrino masses  predicted from Eq.~(\ref{m1second}) are found to be
\begin{align}
m_{1} \simeq (1.1  \to 1.4)  \meV, ~~ 
m_{2} \simeq (8.5 \to 9.1) \, \meV,  ~~ m_{3} \simeq (48 \to 51) \, \meV . 
\label{realmasses}
\end{align}
To show an example, when we set the parameters as follows,
\begin{align}
{m_{1} \over m_{2}} = 0.14 , ~~ \To ~~ U_{e3} = -0.127,
\end{align}
and the numerical value of the $U_{\rm PMNS}$ will be 
\begin{align}
U_{\rm PMNS} = 
\begin{pmatrix}
0.822 & -0.575 & -0.127 \\
0.380 & 0.692 & -0.625 \\
0.455 & 0.466 & 0.772 \\
\end{pmatrix} .
\end{align}
In this matrix, all elements are in $3 \s$ range of Eq.~(\ref{realU}), except $U_{e3}$. 
It shows that the large mixing angles consistent with the experiments are possible from the democratic mass matrices. 
In the second-order perturbation, the predictability becomes little lower. 
However, we can consider  SO(10) GUT or Pati--Salam models for the hierarchical $Y_{\n}$. 
Relating $Y_{\n}$ and $Y_{u}$ in some manner, 
several free parameters in the neutrino sector are expected to be removed. 
Furthermore, the derivation in this paper remains only at the tree level. 
The radiative corrections \cite{Haba:2000rf, Mei:2005gp, Ray:2010rz} and threshold correction \cite{Gupta:2014nba} will modify the results. We leave it for our future work.

\subsection{Relating observables and CP violation}

Effective mass in double beta decay experiment $\vev{m_{ee}}$
\begin{align}
\vev{m_{ee}} = \sum_{i=1}^{3} m_{i} U_{ei}^{2} ,
\end{align}
is calculated from Eqs.~(\ref{UPMNS1}) and (\ref{realmasses}) as
\begin{align}
\vev{m_{ee}} &\simeq  m_{1} \lsp \frac{1}{\sqrt{2}} +  \frac{m_1}{\sqrt{2} m_2} \rsp^{2}
+ m_{2} \lsp -\frac{1}{\sqrt{2}} + \frac{m_1}{\sqrt{2} m_2} \rsp^{2} 
+ m_{3} U_{e3}^{2} \\
& \simeq \frac{m_{2} - m_{1}}{{2}} + m_{3} U_{e3}^{2} 
\, \simeq \, 4.5 \, \meV.
\end{align}
In this study, we assumed all parameters are real. 
Meanwhile, several studies surveys CP violation in democratic matrices \cite{Fukugita:1999xb, Fukuura:1999ze}.
Since the $S_{3}$ symmetry prohibits the relative phase between matrix elements, 
nontrivial phases are associate with the breaking parameters. 
When the breaking term is diagonal, CP violation is introduced by the following replacement
\begin{align}
\Diag{\z_{f}}{\e_{f}}{\d_{f}} ~~ \to ~~ \Diag{\z_{f} e^{i\phi_{1}} }{\e_{f} e^{i\phi_{2}}}{\d_{f} e^{i\phi_{3}}} .
\end{align}
These CP phases can produce baryon asymmetry of universe by the leptogenesis \cite{Fukugita:1986hr}, 
as in the previous study \cite{Fujii:2002jw}.
Furthermore, the leptonic CP phases might be relate the hadronic ones 
in the viewpoint of GUT. In this case, the leptogenesis might also be restricted in some extent by the CKM phase. 

\section{Conclusions and Discussions}

In this paper, we obtain the light neutrino masses and mixings consistent with the experiments, 
in the democratic texture approach. 
The ansatz is that $\n_{Ri}$ are assumed to transform as ``right-handed fields'' $\bf 2_{R} + 1_{R}$ under the $S_{3L} \times S_{3R}$ symmetry.  
The symmetry breaking terms are assumed to be diagonal and hierarchical, 
which is basically same as the previous studies. 
This setup only allows the normal hierarchy of the neutrino masses, 
and excludes both of inverted hierarchical and degenerated neutrinos. 

Although the neutrino sector has nine free parameters, 
several predictions are obtained at the leading order. 
When we neglect the smallest parameters $\z_{\n}$ and $\z_{R}$, 
the resulting neutrino matrix $m_{\n}$ has only three parameters and then determined from 
the neutrino masses $m_{i}$.  
Therefore, all components of the mixing matrix $U_{\rm PMNS}$ are expressed by 
the masses of light neutrinos and charged leptons. 
 From the consistency between predicted and observed $U_{\rm PMNS}$, 
 we obtain the lightest neutrino masses $m_{1}$ = (1.1 $\to$ 1.4)  meV, 
 and the effective mass for the double beta decay $\vev{m_{ee}}\simeq$ 4.5 meV. 
 
 In the second-order perturbation, the predictability becomes little lower. 
However, the hierarchical $Y_{\n}$ can be unified to other Yukawa interactions 
in SO(10) GUT or Pati--Salam models.  
Relating $Y_{\n}$ and $Y_{u}$ in some manner, 
several free parameters in the neutrino sector are expected to be removed. 
Meanwhile, the derivation in this paper remains only at tree level. 
The radiative corrections and threshold corrections will modify the results. We leave it for our future work.

\section*{Acknowledgement}

The author would like to appreciate anonymous referee of PLB for various valuable comments.
This study is financially supported by the Iwanami Fujukai Foundation, and the Sasakawa Scientific Research Grant from The Japan Science Society, No.~28-214.


\end{document}